\documentclass[twocolappendix,numberedappendix,apj]{emulateapj}
\usepackage{natbib}
\usepackage{epstopdf}
\usepackage{epsfig}
\usepackage{txfonts}
\usepackage{mathrsfs}
\usepackage{natbib,caption2}
\usepackage{subfigure}
\usepackage{longtable}
\usepackage{graphicx}
\usepackage{graphics}
\usepackage{keyval}
\usepackage{trig}
\usepackage{float}
\usepackage{calc}
\usepackage{url}
\usepackage{hyperref}
\usepackage[dvipsnames]{color}
\usepackage{lineno}

\def\beq{\begin{equation}}
\def\eeq{\end{equation}}
\def\bey{\begin{eqnarray}}
\def\eey{\end{eqnarray}}
\def\Myr{\, {\rm Myr} }
\def\Gyr{\, {\rm Gyr} }

\def\kpc{\, {\rm kpc} }
\def\msun{M_\odot}
\def\Msun{M_\odot}

\def\kms{\, {\rm km \, s}^{-1} }

\def\Q0{Q_0}

\def\fc{f_{\rm C} }
\begin{document}
\title{Collisions of young disc galaxies in the early universe}
\author{Beibei Guo\altaffilmark{1,2}}
\author{Xufen Wu\altaffilmark{1,2}}
\thanks{Email: xufenwu@ustc.edu.cn}
\author{Guangwen Chen\altaffilmark{1,2}}
\altaffiltext{1}{CAS Key Laboratory for Research in Galaxies and Cosmology, Department of Astronomy, University of Science and Technology of China, Hefei 230026, China; }
\altaffiltext{2}{School of Astronomy and Space Science, University of Science and Technology of China, Hefei 230026, China}



\begin{abstract}
  In the local universe, disc galaxies are generally well evolved and Toomre stable. Their collisions with satellite galaxies naturally produce ring structures, which has been observed and extensively studied. In contrast, at high redshifts, disc galaxies are still developing and clumpy. These young galaxies interact with each other more frequently. However, the products of their collisions remain elusive. Here we systematically study the minor collisions between a clumpy galaxy and a satellite on orbits with different initial conditions, and find a new structure that is different from the local collisional ring galaxies. The clumpness of the target galaxy is fine-tuned by the values of Toomre parameter, $Q$. Interestingly, a thick and knotty ring structure is formed without any sign of a central nucleus in the target galaxy. Our results provide a promising explanation of the empty ring galaxy recently observed in R5519 at redshift $z=2.19$. Moreover, we show that the clumpy state of the collided galaxy exists for a much longer timescale, compared to isolated self-evolved clumpy galaxies that have been widely investigated. 

\end{abstract}


\section {Introduction}
Ring galaxy was first observed by \cite{Zwicky1941}, which is the now famous `Cartwheel' galaxy. More ring galaxies were discovered in the local universe afterwards \citep{arp1966atlas, theys1976ring, conn2011new}. Most of the ring galaxies are the products of head-on collisions between a disc galaxy and a satellite, which are the so called collisional ring galaxies (CRGs) and have been extensively studied \citep[e.g.,][]{lynds1976interpretation,Few_Madore1986,appleton1996collisional}. On the other hand, disc galaxies were found to be more clumpy at redshift of $z \in [1,~5]$ \citep{Conselice+2005}. Those galaxies are composed of a few giant clumps with a mass of approximately $10^9\msun$ and a size of a few kpc for each \citep{Elmegreen_Elmegreen2005}, while there are no spiral arms, bulges nor exponential discs in the clumpy galaxies. It has been proposed that the clumps could be intrinsic structures caused by the gravitational disc instability \citep[e.g.,][]{Noguchi1998,Immeli+2004}.

Galaxies are much closer to each other at higher redshifts, and therefore galaxies merge more frequently in the past, with a merger rate of $(1+z)^{2.3\pm 0.7}$ with $z \in [0.12,~ 0.55]$ \citep{Patton+2002}. Since CRGs are natural products of galaxies collisions, the number density of CRGs is expected to increase with growing redshift. Indeed, observations revealed a $(1+z)^{m}$ with $m=5.2\pm 0.7$ increase of CRGs at $0.1\leq z \leq 1.0$ \citep{Lavery+2004}. The question then arises as to whether the number density of CRGs increases with redshift at $z>1$? Can a galaxy in clumpy state form a CRG undergoing a head-on collision? These problems have been rarely considered in the literature. A systematic study on the gravitational interactions between a clumpy galaxy and a companion galaxy is still lacking.

A recent observation has discovered a massive empty ring galaxy, named as R5519, at z=2.19 \citep{Yuan+2020}, nearby which there is a companion galaxy. \citet{Yuan+2020} believed that this is a CRG in the very young universe. Confusingly, the morphology of this CRG is an symmetric empty ring galaxy without a central nucleus. In previous simulations of the local universe, off-centre collisions are invoked to form the observed empty ring galaxy \citep{Gerber+1992,Mapelli_Mayer2012}. The ring structure observed in the early universe is quite round and thus is more likely to be produced by a head-on collision. However, a head-on collision usually leaves a compact nucleus \citep[e.g.,][]{Gerber+1996}, which was not observed in R5519. Since galaxies are observed to be clumpy at the redshift of the R5519, can a clumpy progenitor undergoing a collision reproduce R5519-like empty ring galaxy?

To understand how gravitational interactions affect the evolution of very young galaxies, we need to simulate the head-on collisions between a clumpy galaxy and a companion galaxy within the framework of the $\Lambda$ cold dark matter (CDM). The disc galaxy is initially Toorme unstable, and evolves rapidly into a clumpy state. We perform serials of disc galaxy models collide with a companion galaxy embedded in a CDM halo. The parameter space of the collisions between galaxies is widely explored through different values of Toomre stability parameter, $Q$, of the disc galaxy model, intruder-to-target-mass-ratio, $M_{\rm i}/M_{\rm t}$ (hereafter short for ITMR), inclination angle, $\theta$, impact parameter, $b$, and concentration of the intruder galaxy. The manuscript is organised as follows: in \S \ref{sec:models}, we begin with the constructions of clumpy galaxy and companion galaxy models for the target and the intruder galaxies, respectively. In \S \ref{sec:nbody}, we will perform the simulations of galaxies collisions with different model parameters. In \S \ref{sec:ts}, we will study the existing timescales of the clumpy state of an isolated and of a collided galaxy, and conclude in \S \ref{sec:conclusion}.

\section{Models}\label{sec:models}
We perform extensive simulations to model the collisions between disc galaxies with different stable conditions and a companion galaxy in the early universe. The disc galaxies are embedded in a spherically symmetric dark matter halo with a truncated isothermal distribution which ensures the flat rotation curves of the disc galaxies in large radii. Disc galaxies are generally gas-rich at the early stage of galaxy evolution. The formation and evolution of clumpy structures caused by Jeans instability of gas has been well established \citep[e.g.,][]{Bournaud+2007,Elemgreen+2008}. Here, we are not intended to revisit the issue of clumps formation. Instead, we wish to study how the clumpy-state galaxies interact with nearby galaxies. Moreover, previous simulations \citep{Theys_Spiegel1976,Gerber+1996,Chen+2018} found a stellar nucleus in the central region of a CRG, irrespective of whether the influence of gas is taken into account or not. This indicates that, the primary driving force of the empty ring structure should be the gravitational interaction, and gas plays only a minor role. We show in Sec. \S \ref{sec:hydro} that the formation and evolution of stellar components in a galaxy with and without gaseous components are very similar. Therefore, we will concentrate on the gravitational evolution of those pre-relaxed clumpy-state galaxies after collisions, which has not yet been explored systematically.

\subsection{A clumpy galaxy as the target galaxy}
The numerical initial conditions (ICs) for target and intruder galaxies are generated from the {\it disk initial conditions environment (DICE)} code \citep{Perret+2014,dice-code}. The ICs are generated by adopting Lagrangian particles. The distribution of these particles are constructed by an algorithm of Metropolis-Hasting Monte-Carlo Markov Chain. The density profile of a galactic disc originally follows an exponential distribution \citep{Hernquist1990}, 
\beq \rho_{\rm d}(R,z)=\frac{M_{\rm d}}{4 \pi h^2 z_0} \exp\left(-\frac{R}{h}\right) {\rm sech}^2\left(\frac{z}{z_0}\right).\eeq
The disc mass is $M_{\rm{d}}$ with a value of $6.0\times 10^{10}~\msun$. The radial scale length, $h$, is $2.0~\kpc$ and all the lengths are scaled in unit of $h$. The disc is originally cut-off at the radius of $5h$. The vertical scale length, $z_0$, is $0.1h$. Cold thin discs are unstable \citep{BT2008}. Different values of Toomre parameter for the stellar and gaseous component \citep{toomre1964}, 
\bey Q_*=\frac{\sigma_{\rm r} \kappa}{3.36 G \Sigma_*},  \nonumber \\
Q_{\rm gas}=\frac{c_s \kappa}{\pi G \Sigma_{\rm gas}}
\eey 
lead to distinguishable disc instability. Here $\sigma_{\rm r}$ and $c_s$ are the radial velocity dispersion of stars and sound speed, $\kappa$ is the epicyclic frequency, $G$ is the gravitation constant, $\Sigma_*$ and $\Sigma_{\rm gas}$ are the surface densities of stars and of gas on the disc. In general, a disc undergoes an instability at low values of Toomre parameter. \citet{Nelson+1998} found that for a threshold of $Q\leq 1.5$, the discs become fragmented in SPH simulations. \citet{Backus_Quinn2016} showed that the disc height is also an important parameter. Thicker discs tend to be more stablised. As a result of such an instability, the formation and evolution of clumps in a young galaxy are different with different initial Toomre parameter, $\Q0$. It is thus important to study how $\Q0$ affects in the interaction of the unstable galaxy with another galaxy. The unstable disc galaxies are started with constant $\Q0 =0.8,~1.0$ and $1.2$, respectively, in all radii of a disc for three disc models in our simulations. Moreover, we setup a galaxy model with $\Q0=1.5$ and relaxed for $2 Gyr$ as a stable target galaxy for a comparison (labeled as ``stable ICs'' hereafter). The results are shown in Sec. \ref{subsec:q}.

The dark matter halo has an NFW profile \citep{Navarro+1997}, which reads
\bey &\rho_{halo}(r) &={\rho_s \over x(1+x)^2}, x=r/r_s,\nonumber \\
&\rho_{s}&={\rho_{cr}\Omega_0\delta_{th}\over 3}{c^3\over \ln (1+c)-{c\over(1+c)}}.\nonumber
\eey
Here $\rho_{cr}=3H_0^2 / 8\pi G$ is the critical density of the Universe, which is determined by the local Hubble constant, $H_0$. The characteristic radius parameter $r_s = 12.5h$. $M_{\rm halo1}=16 M_{\rm{d}}$ is the dark halo mass for the target galaxy. $\Omega_0$ is the fraction of all matter to the critical density. The critical overdensity $\delta_{th}=200$, and therefore the virial mass of the dark halo satisfies a scaling relation of \citep{Mo+1998}
\beq M_{halo1}=\frac{v_c^2r_{200}}{G},\eeq 
where $v_c$ is the circular velocity at the virial radius, $r_{200}$. The value of concentration parameter, $c$, is $4.0$ for the initial conditions at the redshift $z\approx 2.0$ in a study on dark halo profiles by \citet{Bullock+2001}. The numbers of particles in a galaxy model are $1.3\times 10^6$ and $10^6$ for stellar and dark matter component, respectively.  

\subsection{Intruder galaxy}

The intruder galaxy contains a dark matter halo with NFW profile and a stellar component which is described by a Plummer sphere \citep{Plummer1911,BT2008},
\beq
\rho_{\rm plummer}(r)=\frac{3M_{\rm P}}{4\pi r_{\rm P}^3}\left(1+\frac{r^2}{r_{\rm P}^2}\right)^{-5/2},
\eeq
where $M_{\rm P}$ is the total stellar mass of the intruder galaxy and $r_{\rm P}$ is the characteristic scale length of the Plummer profile. The values of $M_{\rm P}$ and $r_{\rm P}$ are $0.5~M_{\rm d}$ and $0.3h$, respectively. The dark halo mass for the intruder galaxy is $M_{\rm halo2}=8~M_{\rm d}$. The overall mass ratio between the intruder and the target galaxy (ITMR), $M_{\rm i}/M_{\rm t}$, is 1:2, originally. The characteristic scale radius for the dark matter of the intruder is $r_s=2h$. The concentration for the NFW halo of the intruder galaxy is $20$. Thus the virial radius of the intruder galaxy is about $80~\kpc$, according to the scaling relation in \citep{Mo+1998}. The numbers of particles are $6\times 10^5$ and $10^5$ for the dark halo and the stellar component, respectively. In the later simulations carried out by tuning the mass-ratios between the intruder and the target galaxies, the parameters for mass model of the intruder galaxy are changed. The new values will be shown in the specific simulations in \S \ref{subsec:imass}.

\section{Gravitational instability and clumpy rings}\label{sec:nbody}
We use an adaptive mesh refinement (AMR) N-body code, ${\it RAMSES}$ \citep{Teyssier2002}, to perform the simulations. The gravitational potential and accelerations of collision-less particles, including stellar and dark matter components, are calculated by means of the particle-mesh scheme. The box size is $500~\kpc$ for the self-evolved simulations, and $2500~\kpc$ for the collision simulations. The minimal and maximal refinement levels are $l_{\rm min}=7$ and $l_{\rm max}=24$, respectively. The maximal refinement level corresponds to a maximal grid resolution of $\Delta x=500\kpc/2^{24}=0.03$pc for the self-evolved models and a maximal grid resolution of $\Delta x = 2500\kpc/2^{24}=0.15$pc for the simulations of collisions. At each coarse time step, the spatial resolution is taken in the range between $l_{\rm min}$ and $l_{\rm max}$. Each AMR cell is further divided into eight sub-cells if there are more than ten stellar or dark matter particles in this cell in a pure N-body simulation. The actual maximal grid resolution reaches to a refinement level of $16$ for the self-evolved models, and of $18$ for the collisional simulations.
The gravitation constant $G=1$ in all the simulations.

In order to examine the impact of gas dynamics, a self-evolution simulation with a gaseous component for an unstable model is provided. The gas dynamics are simulated with a second-order unsplit Godunov method. In a hydrodynamics simulation (\S \ref{sec:hydro}), at each coarse time step, there is a second criterion for grid resolution: the AMR cell is also divided if $\Delta x/\lambda > 0.25$, where $\lambda$ is the Jeans length. By doing so, artificial fragmentation can be avoided \citep{Truelove+1997}. The selection of grid resolution is investigated in more depth in the field of protostellar or protoplanetary disc \citep{Lichtenberg+2015,Deng+2017}. Specifically, the study of self-gravitating protoplanetary discs \citep{Lichtenberg+2015} with adaptive mesh refinement method code, {\it ENZO}, which is similar to {\it RAMSES}, showed that the number of clumps is sensitive to the grid resolution. On the galactic scale, \citet{Tamburello+2015} found that with higher grid resolution, the discs become more stable after a relaxation and the fragmentation is suppressed. We have provided a set of simulations with different grid resolutions to test the gravitational instability amplified by numerical noise in the appendix, \S \ref{sec:resolution}. However, the formation of fragmentation on discs is out of the scope of our current work. Here we are mainly interested in the evolution of clumps (already formed due to fragmentation) after undergoing a collision with a companion galaxy. 
Moreover, the gravitational evolution of the stellar component of the unstable galaxy with and without gas has been compared in \S \ref{sec:hydro}. We find that the evolution of the stellar component is very similar for models with and without gas. Thus it is safe to ignore gas. To focus on the pure gravitational evolution of the clumpy galaxies undergoing a collision, we simplify our models by only considering the stellar and dark matter components. 

\begin{figure}
\centering
\includegraphics[width=85mm]{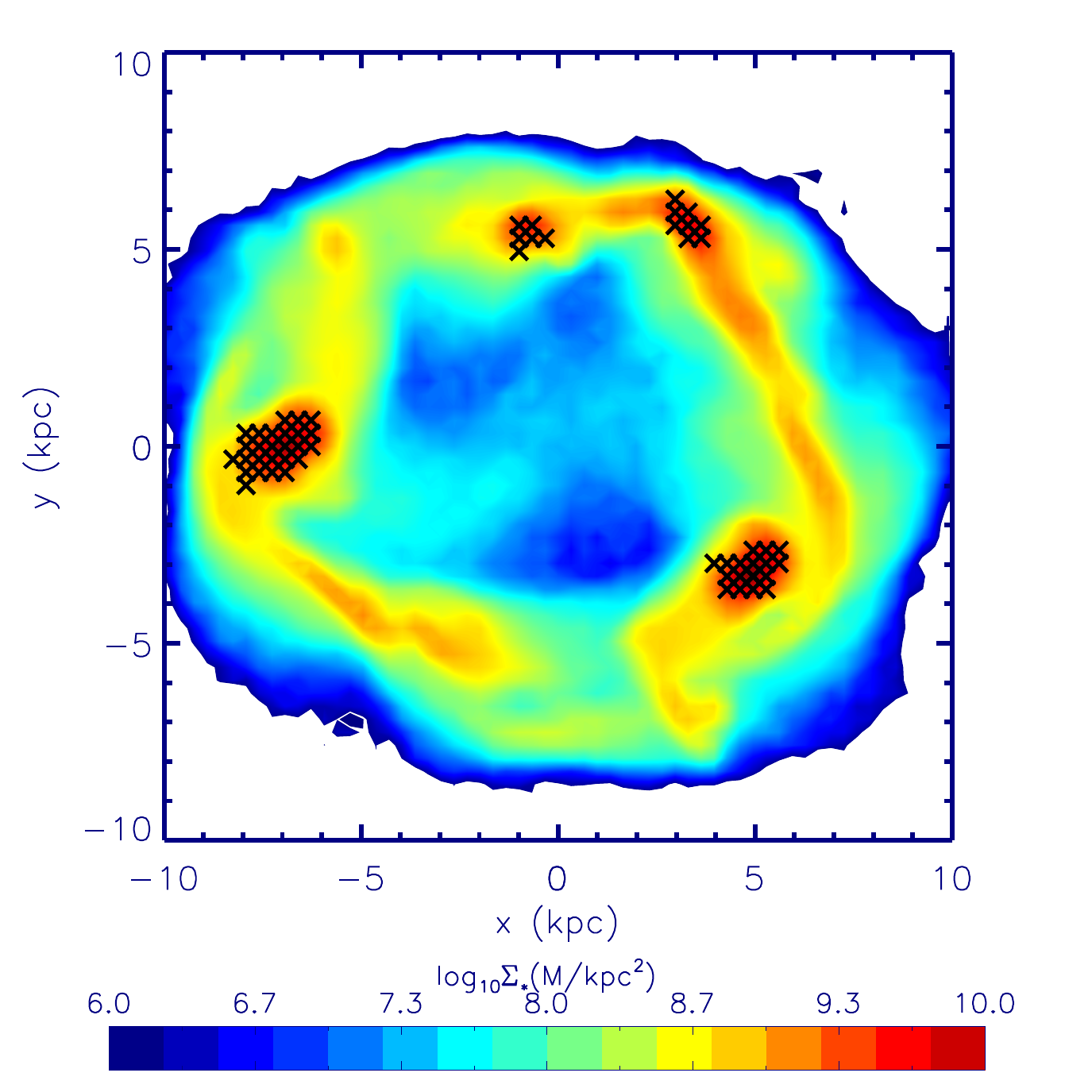}
\caption{Projected mass density profiles of a snapshot (i.e., the upper leftmost panel of Fig. \ref{den2d_q}) for the collisional target galaxy in one simulation in \S \ref{subsec:q}. The regions marked by black crosses are the clumps, which can be distinguished from the collisional ring structures. 
}\label{def-clumps}
\end{figure}

The fixed values of the initial Toomre parameter $\Q0$ allow us to construct gravitational unstable disc galaxy models. To compare the formation and evolution of the clumps with an isolated young galaxy, the models are also self-evolved for 2 Gyr. In the early stage of the self-evolution simulations, the models quickly evolve into a clumpy state. Technically, we define the clumpy structure by the high surface density regions on the galactic plane, i.e., the regions where the surface density, $\Sigma$, is a factor of $3.5$ lager than that of the average surface density in the same radius and the surface density is higher than $5.0 \times 10^8 \msun \kpc^{-2}$. The definition of clumps ensures that the sizes of clumps are smaller than $3~\kpc$, and that clumps can be distinguished from larger galactic structures such as the spiral arms formed due to gravitational instability. Fig. \ref{def-clumps} shows the regions of defined clumps with black crosses for a snapshot of a simulation (the snapshot when a target clumpy galaxy undergoes a collision on $q1$ orbit in Tab. \ref{orbit} and the collisional ring-like structure propagates out to $3h$) in the following section, \S \ref{subsec:q}. The clumps can be distinguished from the collisional ring-like structures, as seen in Fig. \ref{def-clumps}. 

In the self-evolution of the unstable galaxy, the fraction between clumps and overall stellar masses, $\fc$, reaches the maximal value at about $100-200\Myr$, which is consistent with previous studies \citep[e.g.,][]{Bournaud+2007,Elemgreen+2008}. The clumps disappear after $300 \Myr$ in general and all the models approach new stable states at the end of the self-evolution simulations. 

For collisions with an intruder galaxy, the intruder galaxy is moving on a fly-by orbit with an initial position of $(x_0,~y_0,~z_0)$ in a galactocentric coordinate of the disc galaxy, and the initial relative velocity is $(v_{\rm 0x},~v_{\rm 0y},~v_{\rm 0z})$ along the vertical direction of the disc. The model parameters are listed in Table ~\ref{orbit}. We will perform series of simulations including head-on collisions and off-centre collisions between a clumpy galaxy and an intruder galaxy. In the case of head-on collision (\S \ref{subsec:q}), the model is the most simplified. A target galaxy generated by different values of Toomre instability parameter collides with an companion galaxy. Several sets of off-centre simulations will be performed later on more generic orbits. Those orbits contain more varying parameters, including different values of inclination angle $\theta$ (\S \ref{subsec:incl}), initial relative velocity $v_0$ (\S \ref{subsec:v0}), impact factor $b$ (\S \ref{subsec:b}). Moreover, the collisional structures of the clumpy target will be studied by tuning the mass ratios between the intruder and the clumpy target galaxies (\S \ref{subsec:imass}). 

\begin{table*}
\centering
\caption{
The model parameters of the collisions between the unstable target galaxy and the intruder galaxy. The names of the orbits, ITMR and the Toomre instability parameter ($Q_0$) of the target galaxies are given in the $1^{st}-3^{rd}$ columns. The initial positions of the intruder galaxy in galactocentric coordinates of the targets are shown in the $4^{th}$ column. The $5^{th}$ column lists the impact parameter, $(b)$. The initial relative velocity, $v_{\rm 0}$, and its x$-$ and z$-$axial components, $v_{\rm 0x}$ and $v_{\rm 0z}$ are shown in the $6^{th}-8^{th}$ columns. The $9^{th}$ column is the inclination angle, $\theta$. The $10^{th}$ through $12^{10}$ columns represent the timescales of the collisions and the masses of the rings when they propagate out to $3h$ and $4h$.
}\vskip 0.5cm
\begin{tabular}{llllllllllll}
\hline
Collisional  & $M_{\rm i}/M_{\rm t}$ & $Q_0$& $(x_0,~y_0,z_0)$ & $b$ &$v_{\rm 0}$ & $v_{\rm 0x}$ & $v_{\rm 0z}$ & $\theta$ & $t_{\rm{coll}}$  &$M_{\rm{ring,~3h}}$&$M_{\rm{ring,~4h}}$ \\
models & &  & (kpc) & (kpc) & (km/s) & (km/s) & (km/s) & ($^\circ$) & (Myr)  & ($10^{10}\Msun$)& ($10^{10}\Msun$)\\
\hline
q0 &0.5 & 1.5 & $(0.0,~0.0,~-50.0)$ &0.0&600 &0& 600 & 0.0 & 68.0 &3.6 & 2.2 \\
q1 &0.5& 0.8 & $(0.0,~0.0,~-50.0)$ &0.0&600& 0 & 600 & 0.0 & 70.9 &5.5 & 4.5 \\
q2 &0.5& 1.0 & $(0.0,~0.0,~-50.0)$ &0.0&600& 0 & 600 & 0.0 & 71.0 &5.7 & 4.3 \\
q3 &0.5& 1.2 & $(0.0,~0.0,~-50.0)$ &0.0&600& 0 & 600 & 0.0 & 71.0 &5.7 & 4.8 \\
t1 &0.5& 0.8 & $(0.0,~0.0,~-50.0)$ &0.0 & 650& 0 & 650 & 0.0 & 62.9 & 5.4 &-\\
t2 &0.5& 0.8 & $(-28.9,~0.0,~-50.0)$&0.0 & 650& 325 & 563 & 30.0& 73.1& 5.4 &-\\
t3 &0.5& 0.8 & $(-50.0,~0.0,~-50.0)$&0.0 & 650& 460 & 460 & 45.0& 91.1& 4.7 &-\\
v1 &0.5& 0.8 & $(-28.9,~0.0,~-50.0)$&0.0 & 550& 275 & 476 & 30.0& 83.7& 4.6 &-\\
v2 &0.5& 0.8 & $(-28.9,~0.0,~-50.0)$&0.0 & 750& 375 & 650 & 30.0& 63.4& 5.7 &-\\
b1 &0.5& 0.8 & $(-31.2,~0.0,~-50.0)$&2.0 & 650& 325 & 563 & 30.0& 73.3& 3.1 &-\\
b2 &0.5& 0.8 & $(-33.5,~0.0,~-50.0)$&4.0 & 650& 325 & 563 & 30.0& 73.5& 1.4 &-\\
m1 &0.1& 0.8 & $(0.0,~0.0,~-50.0)$ &0.0 &650 &0& 650 & 0.0 & 68.9& 2.8 &-\\ 
m2 &0.1& 0.8 & $(-28.9,~0.0,~-50.0)$ &0.0 &650 &325& 563 & 30.0 & 75.3& 2.8&-\\
m3 &0.1& 0.8 & $(-50.0,~0.0,~-50.0)$ &0.0 & 650 &460 & 460 & 45.0& 83.3& 2.2&-\\
m4 &0.3& 0.8 & $(0.0,~0.0,~-50.0)$ &0.0 &650 &0& 650 & 0.0 & 67.7 & 5.2 &-\\
m5 &0.3& 0.8 & $(-28.9,~0.0,~-50.0)$ &0.0 &650 &325& 563 & 30.0 & 72.3 & 4.3&-\\
m6 &0.3& 0.8 &$(-50.0,~0.0,~-50.0)$ &0.0 & 650 &460& 460 & 45.0 & 93.4 & 3.9&-\\
%
\hline
\end{tabular}
 \label{orbit}
\end{table*}

\subsection{Head-on collisions between a clumpy target galaxy and an intruder galaxy}\label{subsec:q}

Firstly, a set of head-on collisions between a target galaxy with different values of $\Q0$ and an intruder galaxy has been performed with model parameters named $q1$-$q3$. As aforementioned, we take a stable model, i.e., the one with $\Q0 = 1.5$ and freely relaxed for $2~\Gyr$, as a progenitor to collide with the dwarf galaxy on the same orbit for a comparison, which is the $q0$ model. The collisional event happens at a timescale of approximately $70~\Myr$ on such an orbit, at the time when clumps are forming due to gravitational instability. 

\begin{figure*}
\centering
\includegraphics[width=170mm]{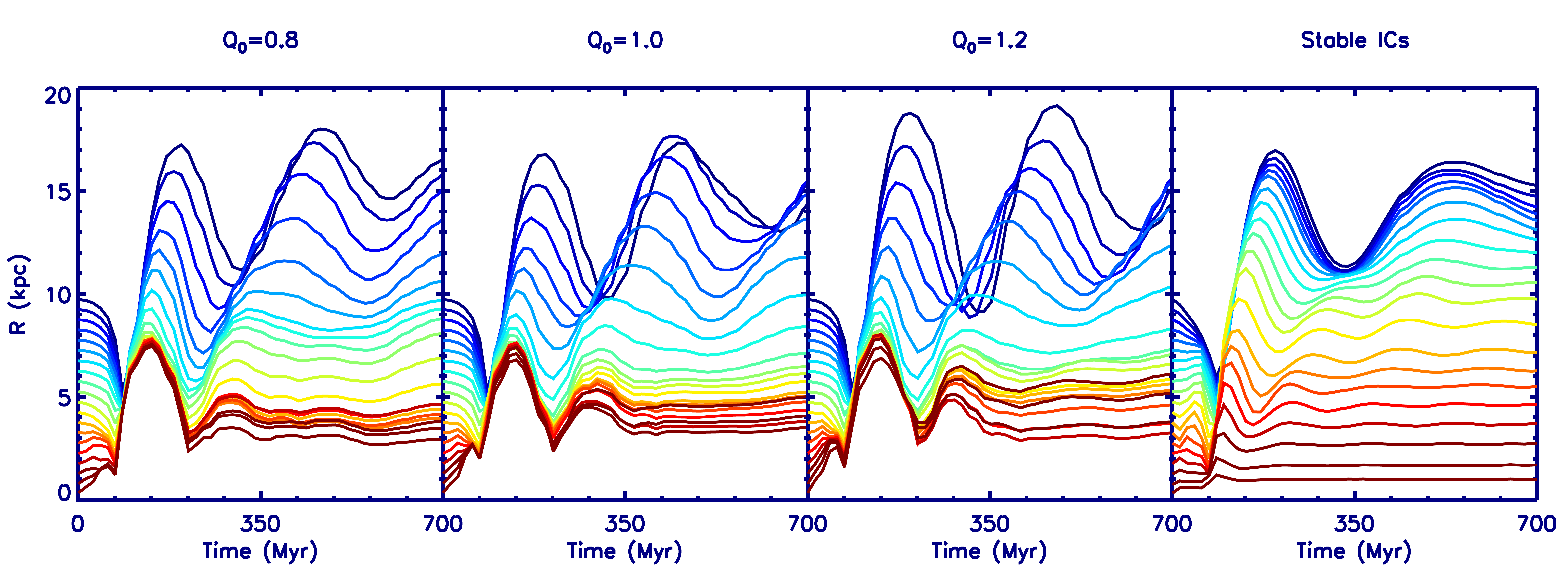}
\caption{The temporal evolution of mean face-on radii of the stellar particles within $0.25h$ radial intervals for the ICs of the $q1-q3$ models (the leftmost to the third panel) and of the $q0$ model (the rightmost panel).}\label{temporal}
\end{figure*} 

\begin{figure*}
\centering
\includegraphics[width=150mm]{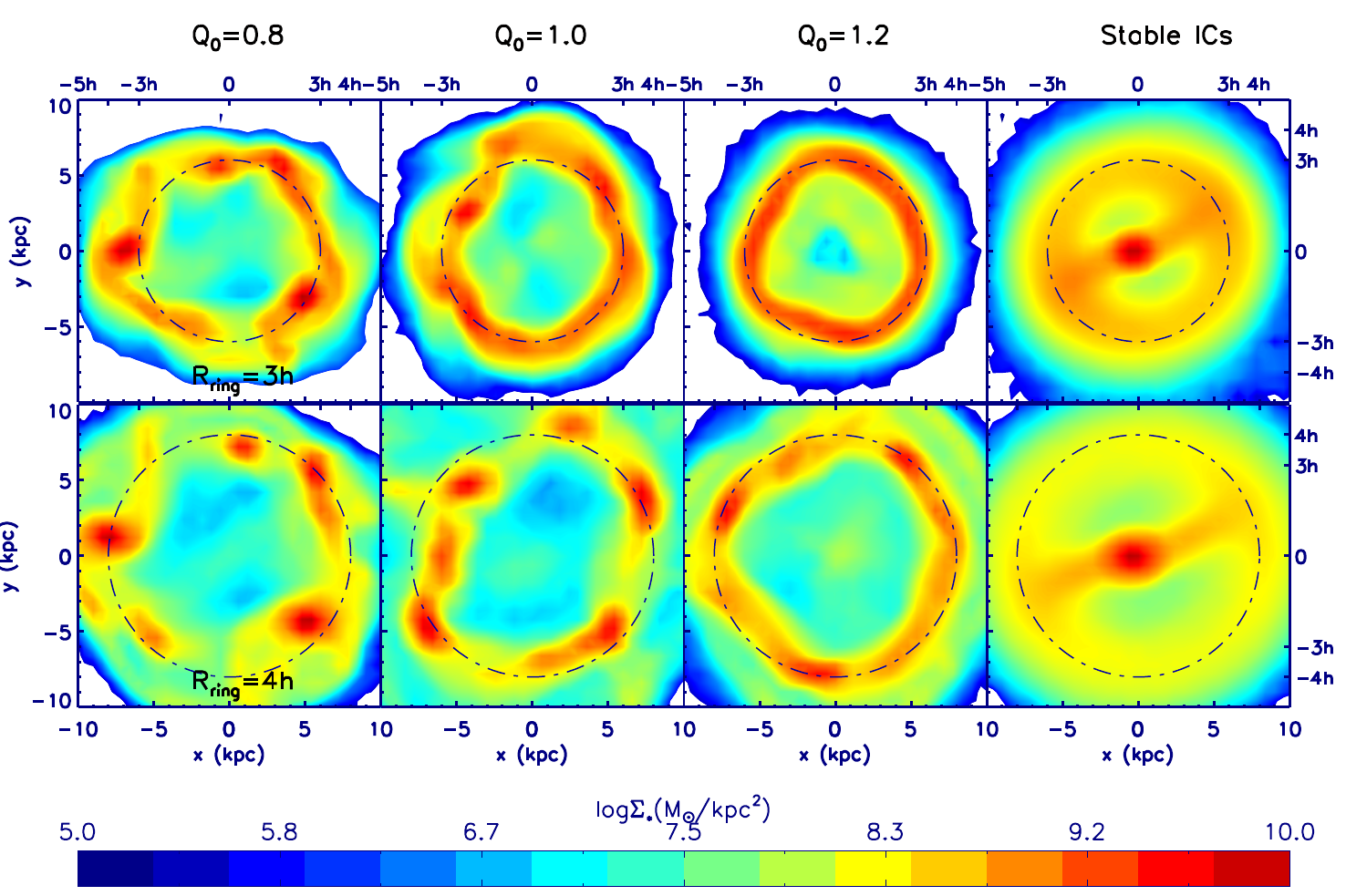}
\caption{Projected mass density profiles of the collisional rings propagating to the radii of 3h (upper panels, dotted-dashed circles) and of 4h (lower panels) of the target galaxy for models $q0$-$q3$. The values of Toomre parameter, $Q$, are taken as constants for each model. $\Q0 ~=~0.8,~1.0$ and $1.2$, individually, for the varying $\Q0$ models ($q1$-$q3$ in Tab. \ref{orbit}). The rightmost panels are the CRGs form from stable initial conditions (ICs) for a comparison (model $q0$ in Tab. \ref{orbit}).}\label{den2d_q}
\end{figure*} 

The temporal evolution of the mean radii of stellar particles for models $q0$-$q3$ are shown in Fig. \ref{temporal}. Initially, at $T=0$, the stellar particles in a model are binned radially within each $0.25h$ radial intervals between $0$ and $5h$ on the projected disc plane. The curves in Fig. \ref{temporal} present the temporal evolution of the mean radii of these initially binned particles within $700~\Myr$. For the $q1$-$q3$ models, the mean radii of the inner particles for the ICs expand outwards to approximately $8~\kpc$, i.e., $4h$, after the collisions. It implies that the central regions of the $q1$-$q3$ models are almost empty at $T\approx 100\Myr$. At $T\approx 300~\Myr$, the mean radii of the innermost binned particles are around $2.5h$, indicating the formation of the second ring. The propagation radii of the second rings in all the models are smaller than the first rings. For the stable model (the rightmost panel of Fig. \ref{temporal}), the mean radii of the inner bins are much smaller than those of the unstable models.

\subsubsection{The shapes of the collisional structures}
\begin{figure}
\includegraphics[width=90mm]{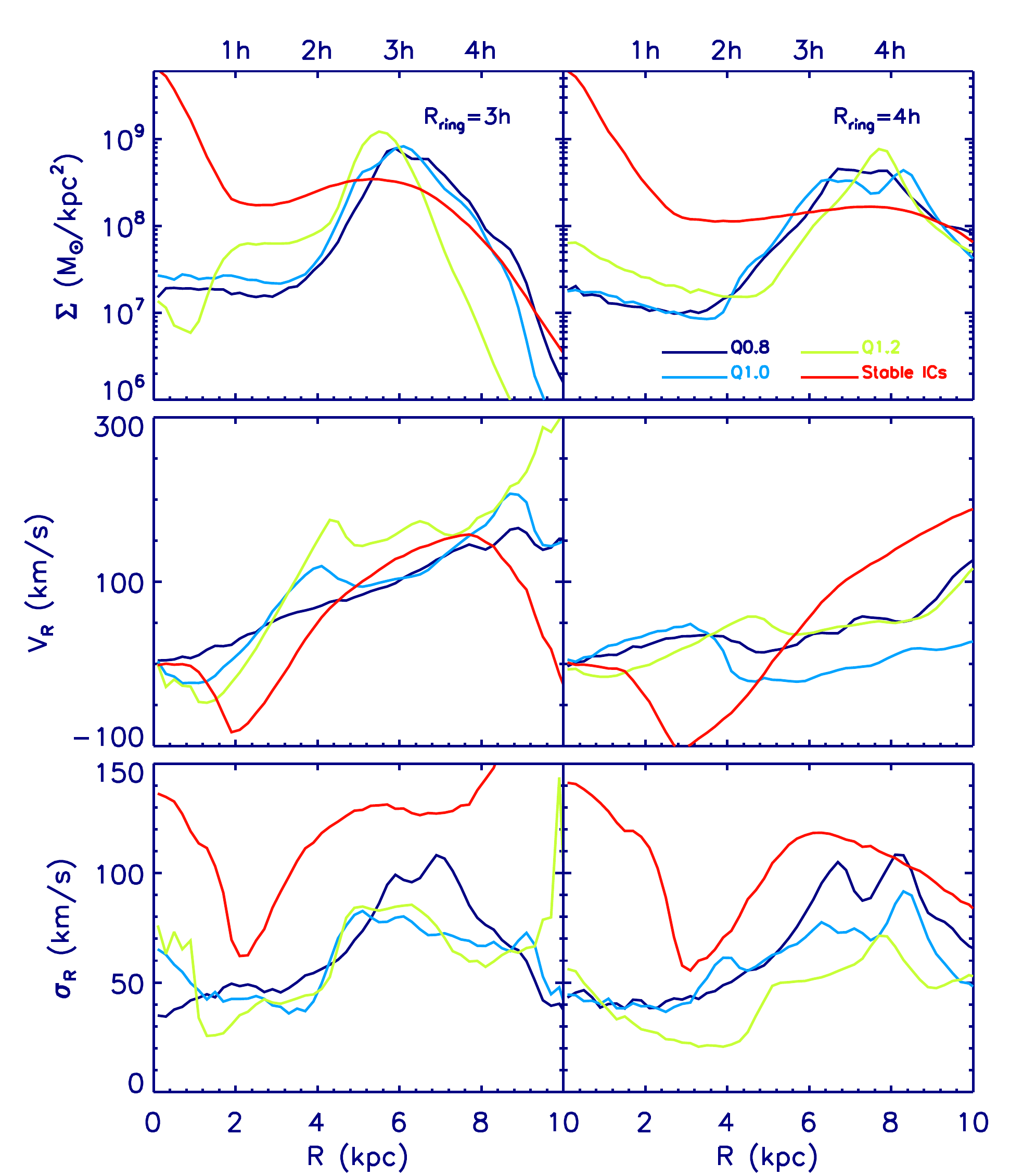}
\caption{Projected mass density, $\Sigma(R)$ (upper panels), radial velocity, $v_R(R)$ (middle panels) and radial velocity dispersion, $\sigma_R(R)$ (lower panels) profiles of the collisional rings propagating to the radii of 3h (solid curves in left panels) and of 4h (solid curves in right panels) of the target galaxies. Different line colours indicate different values of $\Q0$ for the clumpy progenitors and the stable ICs. }\label{den1d_q}
\end{figure}

After head-on collisions (models $q0$-$q3$), ring(-like) structures form on the disc planes and propagate radially outwards in all the four target models with different values of $\Q0$ and in the stable target galaxy. The projected mass densities of the target galaxies on their disc planes are presented in Fig. \ref{den2d_q} when the whole ring structures propagate outwards to $3h$ (upper panels) and $4h$ (lower panels), which is approximately $30~\Myr$ and $50~\Myr$ after the collisions, respectively. The target galaxies exhibit different behaviours. The rings tend to be more knotty for progenitors with smaller values of $\Q0$. This means that the clumps spread outwards associated with the propagation of the rings. Since the radii of the clumps in a ring in models $q1$-$q3$ are slightly different, the massive clumps lead to multiple density peaks in the projected densities profiles(upper panels of Fig.\ref{den1d_q}). With the increase of $\Q0$, the ring structures become smoother. For model $q3$, the ring appears to be smooth at $3h$. However, when the ring propagates to $4h$, the knotty structures can still be found on the projected density.

Moreover, no central nuclei are found in the unstable models. Interestingly, the mass densities of the centres are significantly lower than that of the ring for all of the clumpy models (models $q1$-$q3$). The empty central regions can be readily seen in the 1-dimensional projected density figures (solid curves in upper panels of Figure \ref{den1d_q}), in which the central projected densities of the CRGs formed from unstable progenitors are significantly lower than that from stable ICs (model $q0$). The ring of the stable ICs model is weaker than other models, since this model contains a massive bulge fully relaxed from the $\Q0=1.5$ model, which obviously stablises the disc.

The formation of empty collisional ring galaxies is quite different from previous relevant works \citep[e.g.,][]{Lynds_Toomre1976,Theys_Spiegel1976,Smith+2012,Chen+2018}, which always found a central stellar nucleus in a simulated head-on or off-centre CRG. In these previous simulations, there are always compact nuclei left in the stellar components of the target galaxies. In our comparison example (the rightmost panels in Fig. \ref{den2d_q}), compact central regions also exist in the remnants of collisions. In reality, empty ring galaxies have been observed in the local universe, such as Arp 147. To explain the formation of the local empty ring galaxies, off-centre collisions are requires. \citet{Gerber+1992} proposed that a projection effect to the observer on the earth accounts for the absence of central nucleus of Arp 147, while \citet{Mapelli_Mayer2012} found that the nucleus can be off-set from the centre and be entirely buried within the ring in the off-centre collisions with a narrow range of impact parameter, $b$ ($b\approx 2.1h$ or $3.2h$ and $h=3.7$ kpc in their models).

Here we have provided another natural explanation for the formation of the observed empty and knotty ring galaxy R5519 \citet{Yuan+2020}. An advantage of our mechanism is that we do not need to introduce any special projection angle nor a narrow range of the impact parameter for an off-centre collision. We have shown that a head-on collision between a clumpy progenitor and a dwarf galaxy can result in an empty ring, which is not expected in studies of the local CRGs \citep[e.g.,][]{Lynds_Toomre1976,Theys_Spiegel1976,Smith+2012,Chen+2018,Gerber+1992,Mapelli_Mayer2012}. However, the hypothesis of head-on collision imposes a very special condition on the interactions of galaxies. A natural question that arises here is whether an empty ring galaxy form in a more general orbit when two galaxies collide with each other. The collisional structure formed through off-centre collisions will be systematically studied in the following subsections.

\begin{figure*}
\includegraphics[width=160mm]{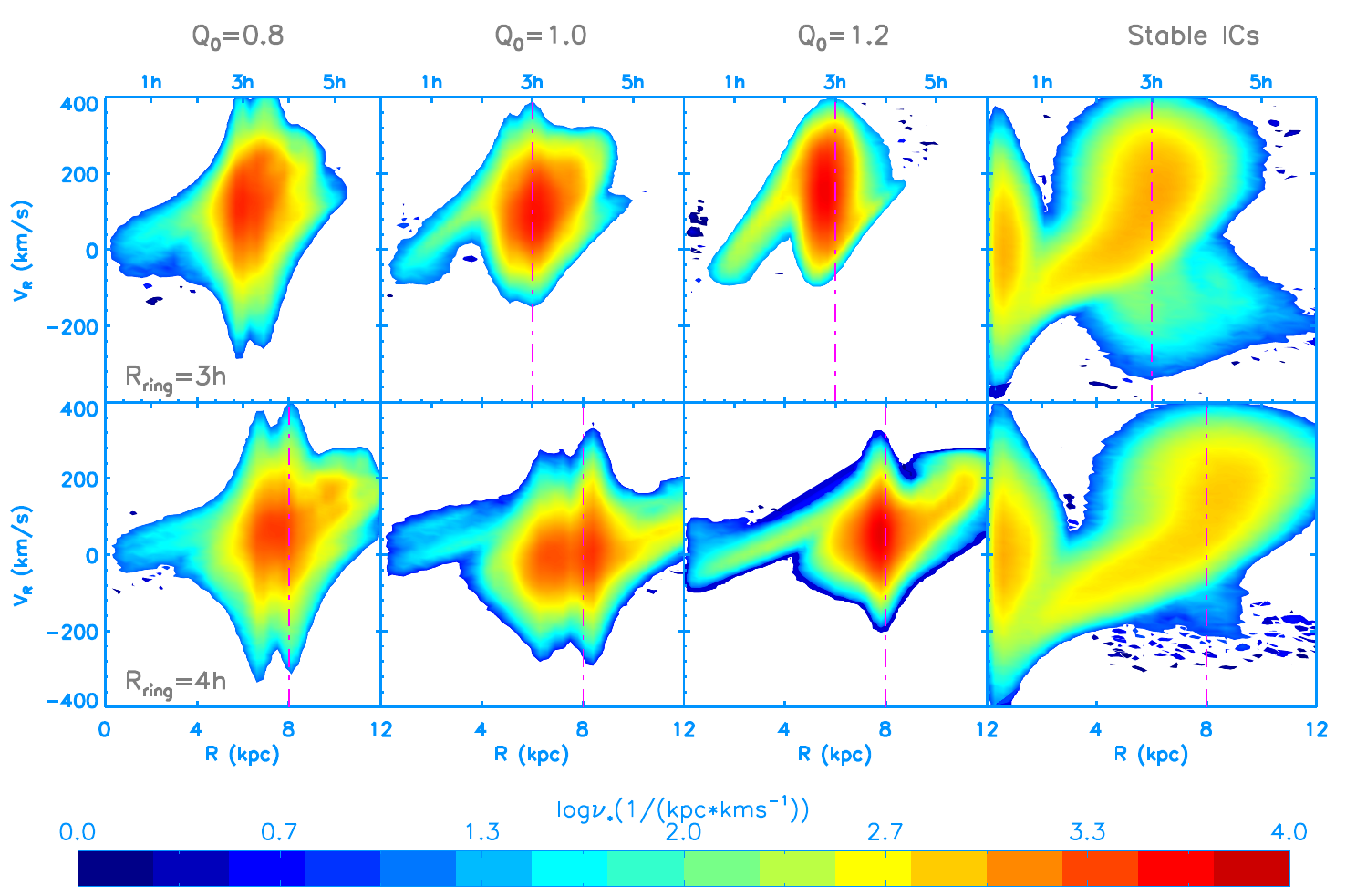}
\caption{The phase-space distribution of the stellar components for the target galaxies when their rings propagate out to $3h$ (upper panels) and $4h$ (lower panels), which are indicated with red dotted-dashed lines. The colourbar shows the number density of the numerical particles in the phase-space distribution.}
\label{psds}
\end{figure*}

\subsubsection{The mass of the ring-like structures}
Fig. \ref{den2d_q} and the upper panels of Fig. \ref{den1d_q} show that a large fraction of stellar particles are distributed near the ring-like structures in the clumpy target galaxies. There are no central compact nuclei in these models. We examine the mass of the rings when they propagate out to $3h$ and $4h$ (Tab. \ref{orbit}) for models $q0$-$q3$. The rings are defined at their peak values of 1-dimensional face-on projected density, $\Sigma$, and the thickness of a ring is $\pm h$ around the peak. For the unstable models, when the rings are at $3h$, the mass of the rings are  over $5.5\times 10^{10}\msun$, which corresponds to a fraction of $91.7\%$ of the stellar mass. When the rings propagate out to $4h$, the masses of the rings are $4.3\times 10^{10}$-$4.8\times 10^{10}\msun$. This is a fraction of over $70\%$ of the stellar mass in a galaxy. Thus the majority of the disc mass is distributed in the knotty ring in a unstable model. This is different to that of a stable model. In $q0$ model, the masses of the rings are $3.6\times 10^{10}\msun$ and $2.2\times 10^{10}\msun$ at $3h$ and $4h$, respectively. The mass fraction of the ring is much smaller in a stable model, as there is a compact central nucleus. When the galaxies collides each other, the particles in the nucleus do not propagate outwards, and the ring particles are basically from the disc particles \citep{Chen+2018}.


\subsubsection{Kinematics of the rings}
When the rings propagates out to $R=3h$ (middle left panel of Figure \ref{den1d_q}), the radial velocity profiles, $v_R(R)$ for the unstable models with $0.8\leq \Q0 \leq 1.2$, keep increasing with the growth of radius. Clearly, the stars move on radial orbits on the disc plane, especially near the position of the rings. There are multiple peaks for the $v_R(R)$ profiles around the position of the thick rings, i.e., between $2h$ and $4h$, which is related to the radial motion of the clumps. The $v_R(R)$ profile of the stable model has a sharper peak at the position of approximately $4h$. However, when the rings propagates out to $4h$ (middle right panel of Figure \ref{den1d_q}), the $v_R(R)$ for the unstable models are almost flat and near zero. This implies that the rings are approaching the maximal propagation radius. In the later stage, the rings are going to collapse backwards the galactic centre. This is different from that of the stable model, in which the  $v_R(R)$ profiles are still arising with the increase of radius when the ring is at $4h$. A ring can propagate to a further distance to the galactic centre in a stable model. However, the contrast of the surface density profile $\Sigma(R)$ of the stable model around $4h$ appears to be weak. The $\Sigma(R)$ and $v_R(R)$ profiles imply that the ring is still moving outwards but the structure is vanishing.

The velocity dispersion profiles, $\sigma_R(R)$, of the collisional ring galaxies are shown in the lower panels of Fig. \ref{den1d_q}. Around the radii of the rings, there are multiple peaks in the $\sigma_R(R)$ profiles of the clumpy galaxies, indicating that the ring structures are dynamically heated by the head-on collision. The $\sigma_R(R)$ profile for the stable ICs model behaves quite differently. There is a peak at the central region, which is consistent with the density peak in the centre of the galaxy model, i.e., the central nucleus. In addition, the $\sigma_R(R)$ profile for the stable model is smoother at the radii of the rings.

\subsubsection{Phase-space distribution of the targets}

The surface density distribution and the radial velocity profiles imply that the phase-space density (PSD) distributions for the clumpy CRGs are quite different from that for the stable CRG model, which is shown in Fig. \ref{psds}. The phase densities of all models when rings propagate out to $3h$ (upper panels) and $4h$ (lower panels) are binned in $0.3~\kpc$ space interval and in $8.0~\kms$ velocity interval. At the regions of the rings, the PSD distributions appear to be the highest. Compared to the unstable models, several different behaviours are observed in the stable model. Apart from a dense region near the position of a ring, the core area also exhibits a high density region. The PSDs for the stable ICs model are lower then the $q1$-$q3$ models. The reason is that there is a central compact nucleus and there are fewer particles in the ring structures at both $3h$ and $4h$. Moreover, the $v_{\rm R}(R)$ of the particles between the galactic centres to the rings are positive for the unstable models, implying that particles inside the rings are moving outwards the galaxies. However, the $v_{\rm R}(R)$ of the particles between the core region and the ring of the stable model are negative and most of the stellar particles in these radii are moving inwards.

\subsection{Off-centre collisions}
\begin{figure}
\centering
\includegraphics[width=90mm]{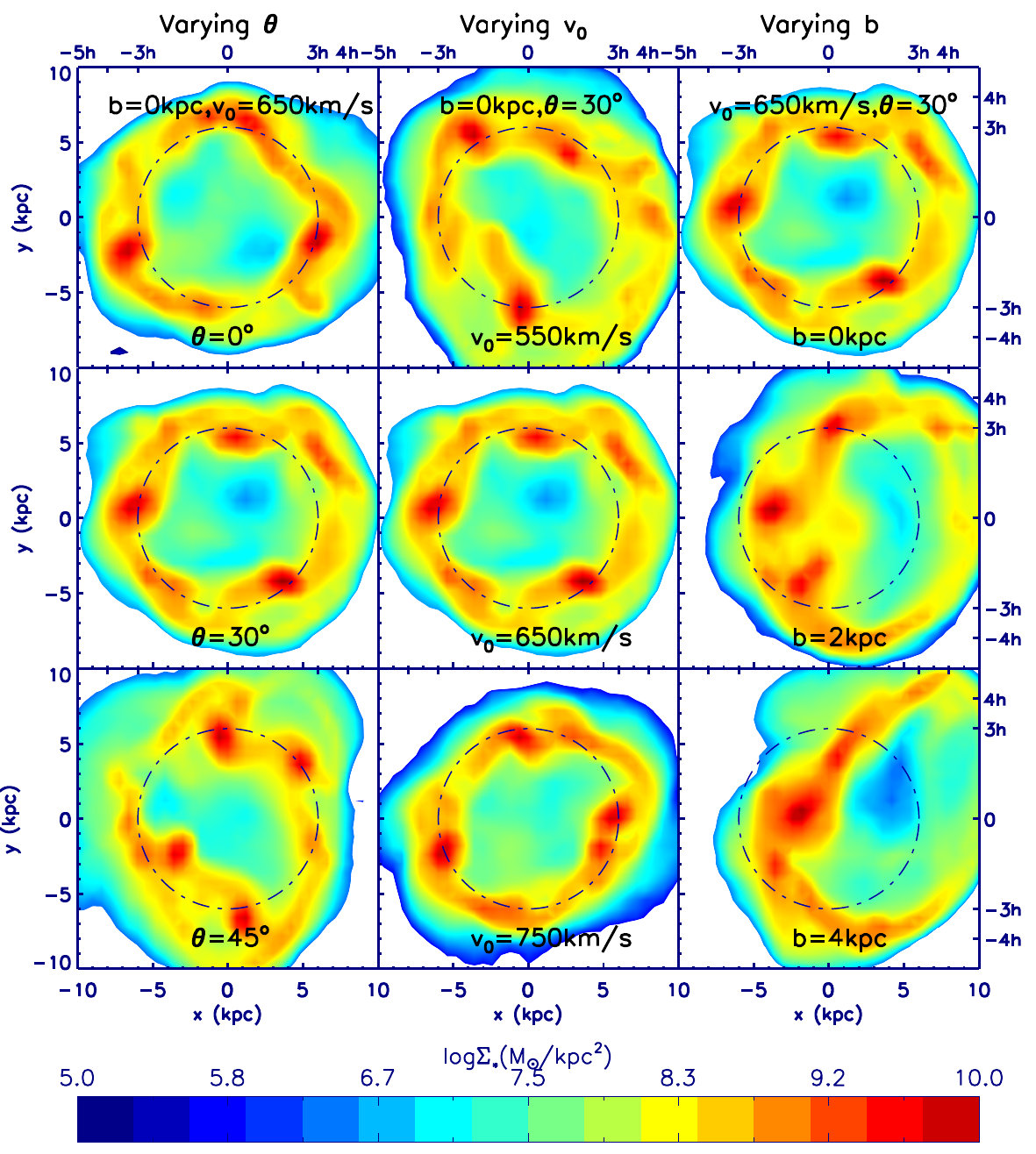}
\caption{Projected mass density profiles of the collisional rings propagating to the radii of 3h. The two galaxies collide on off-centre orbits with different values of $\theta$ (i.e., models $t1-t3$ in Tab. \ref{orbit}, left panels), $v_{\rm 0}$ (i.e., models $v1$, $t2$ and $v2$, middle panels) and $b$ (i.e., models $t2,~b1$ and $b2$, right panels).}\label{den2d_tvb}
\end{figure} 

\begin{figure}
\centering
\includegraphics[width=90mm]{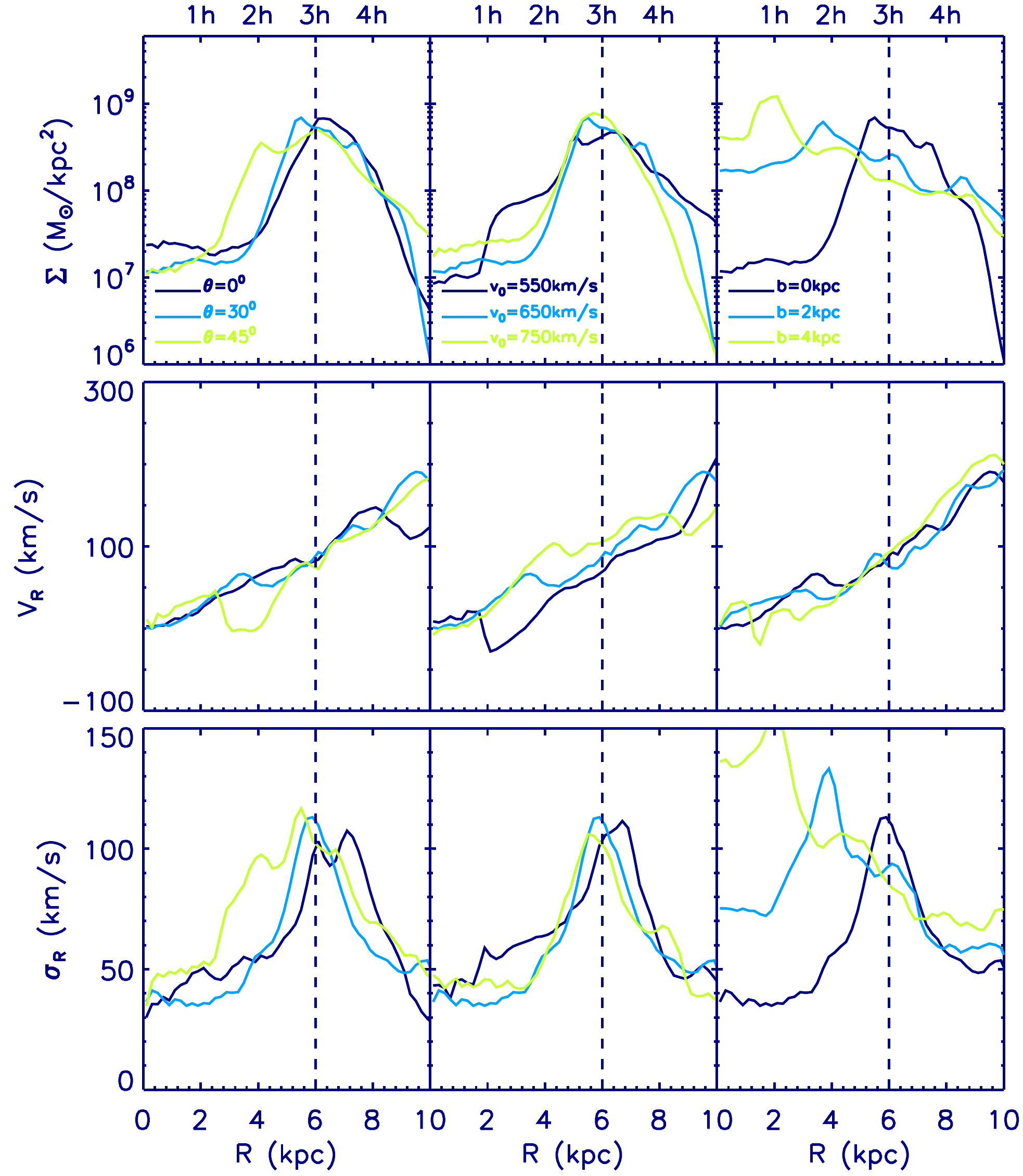}
\caption{Projected mass density, $\Sigma(R)$ (upper panels), radial velocity, $v_R(R)$ (middle panels) and radial velocity dispersion, $\sigma_R(R)$ (lower panels) profiles of the collisional rings propagating to the radii of $3h$ for the target galaxies on orbits with varying $\theta$ (left panels), $v_{\rm 0}$ (middle panels) and $b$ (right panels). }\label{den1d_tvb}
\end{figure}

\begin{figure}
\centering
\includegraphics[width=90mm]{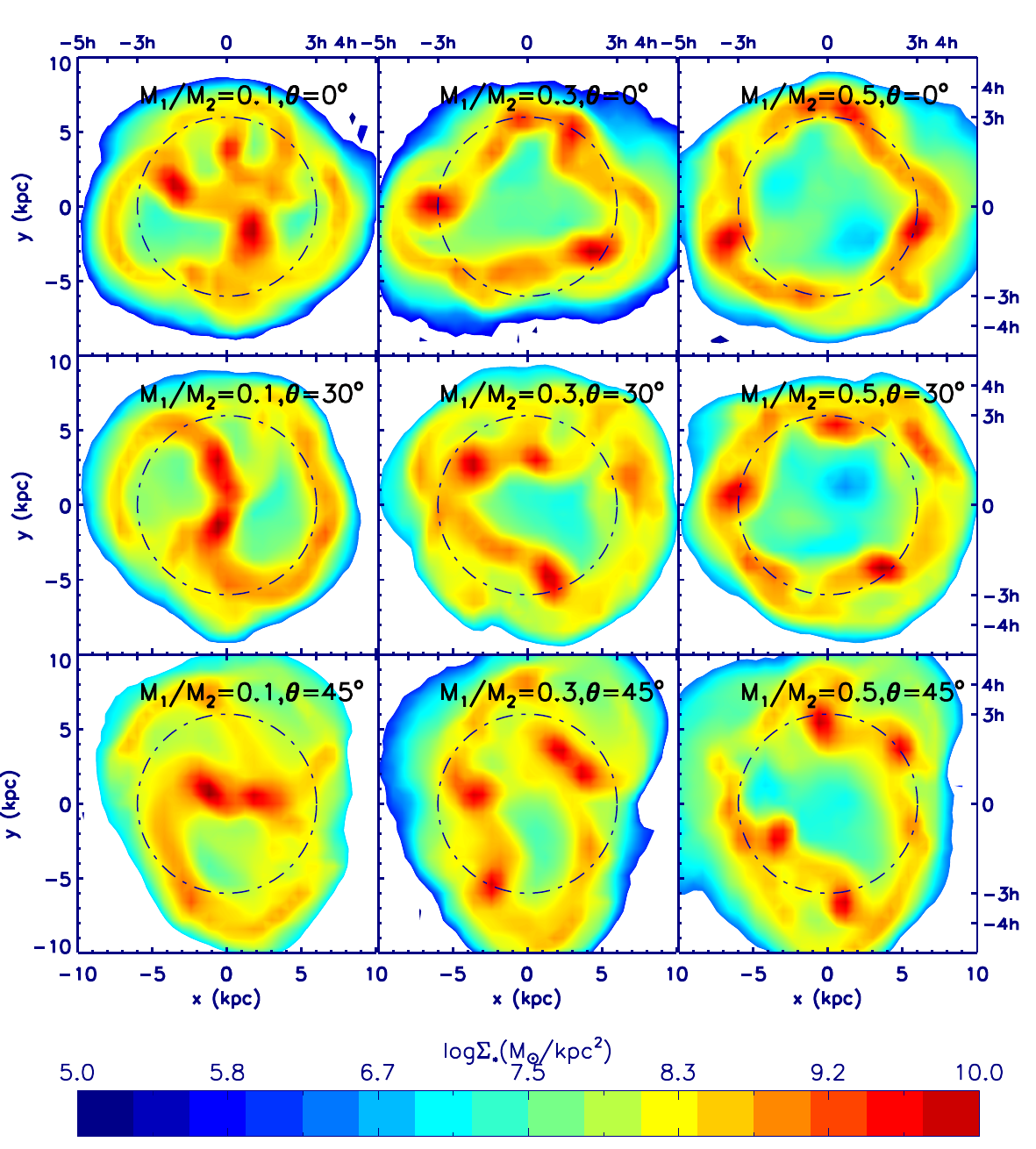}
\caption{Projected mass density profiles of the collisional rings propagating to the radii of 3h. The two galaxies collide on orbits with different values of ITMR (from left to right panels) and with different values of $\theta$ (from upper to bottom panels)}\label{den2d_mt}
\end{figure} 

We perform a set of simulations to explore the ring structures formed through collisions on more generic orbits. The parameters include inclination angle, $\theta$ (models $t1$-$t3$ in Tab. \ref{orbit}), initial relative relative velocity, $v_{\rm 0}$ (models $v1$, $v2$ and $t2$), and impact factor, $b$ (models $t2$, $b1$ and $b2$). The above parameters are defined as follows,
  \bey \theta & = &\tan^{-1}\left(\frac{v_{0x}}{v_{0z}}\right),\nonumber \\
  z_0 &= & {\rm sign}(x_0)\frac{v_{z0}}{v_{0x}}(|x_0|-l),\nonumber \\
  b &= & l\cos\theta,
  \eey
where $l$ is the distance between the original point and the intersection of $\vec v_0$ and the plane of disc. The schematic diagram for the off-centre orbits is the same as that in Fig. 3 of \citet{Chen+2018}.\footnote{There is an error in the last row of Eqs. 7 in \citet{Chen+2018}. The impact factor, $b$, should be $l\cos\theta$.} The 2-dimensional projected density of the target galaxy models at the timescale when the rings propagate out to $3h$ are presented in Fig. \ref{den2d_tvb}.

\subsubsection{Tuning inclination angles}\label{subsec:incl}
In models $t1$-$t3$, the galaxy pairs are on orbits with different values of inclination angle, $\theta$. The inclination angle are tuned by choosing the initial positions of the intruder galaxy. All the models form a knotty ring structure after collisions. When $\theta$ increases, the rings become more elliptical. In the 1-dimensional projected density profile (left upper panel of Fig. \ref{den1d_tvb}), there are multiple peaks near the position of the ring for models $t2$ and $t3$. The multiple peaks are caused by the different radii of the clumps in the elliptical rings. The centres of mass (CoMs) of the ring galaxies for models $t1$-$t3$ are in the geometrical centres of the galaxies, as the tuning parameter is inclination angle for these models. The two galaxies collide each other through their geometrical centres. 

The $v_{\rm R}(R)$ profiles are increasing the growth of the radius, which is similar to those of head-on collisional models, $q1$-$q3$. There are multiple peaks in the velocity dispersion profiles, $\sigma(R)$, which are caused by the different radii of the clumps in the elliptical ring structure. 

\subsubsection{Tuning initial relative velocities}\label{subsec:v0}
To study how the initial relative velocity, $v_{\rm 0}$, change the results of CRGs, we perform a set of simulations with different values of $v_{\rm 0}$. The galaxy pairs are on orbits with a fixed inclination angle of $30^\circ$ and an impact parameter of zero. The collisional rings are knotty and elliptical on these orbits. The rings appear to be similar in these models. The slight difference is attributed to the difference in the strength of gravitational perturbation introduced by the intruder galaxy. The models with larger initial relative velocities generate clearer ring structures. In contrast, the $v1$ model with the smallest $v_{\rm 0}$ forms a fainter ring. The contrast of $\Sigma(R)$ profile for $v1$ model is the lowest in the middle panel of Fig. \ref{den1d_tvb}. In addition, in models with larger values of $v_{\rm 0}$, the rings are more massive, which agrees well with the clearer ring morphology.

\subsubsection{Tuning impact factor}\label{subsec:b}
In order to further study how a clumpy galaxy evolves after colliding with another galaxy on an arbitrary orbit, we set up a set of orbits with a non-zero inclination angle and with tuning impact parameter, $b\in [0, 2h]$ (right panels of Figs. \ref{den2d_tvb} and \ref{den1d_tvb}). The target galaxies manifest a ring-like/arc-like structure after collisions. As the impact factor increases, the ring becomes more lopsided. When the impact factor $b$ takes its largest value, $b=2h$ in the $b2$ model, the ring becomes sufficiently lopsided such that the original closed ring structure is replaced with an arc.
Thus the impact factor cannot be too large to form a closed collisional ring. The peaks of $\Sigma(R)$ and $\sigma(R)$ profiles for model $b1$ and $b2$ do not overlap with the positions of $3h$, due to the lopsidedness of the ring or the arc. It is difficult to define the positions of rings or arcs. Therefore we choose the same timescale when a ring propagates out to $3h$ for the $t2$ model. At the timescale, the right hand sides of the ring or of the arc have already propagate beyond $3h$ for $b1$ and $b2$ models. But the left hand sided of the structures still have not reached $3h$.

\subsection{Collisions with different ITMR}\label{subsec:imass}
The critical ITMR is $1:10$ in the previous studies on the collisional ring galaxies \citep{Hernquist_weil1993}. If the ITMR is smaller than $1:10$, no collisional rings are formed through such weak gravitational perturbations in the stable disc models. Here we also study the galaxy collisions with tuning ITMR. The mass ratios are turned down in models $m1$-$m6$ in Tab. \ref{orbit}. In models $m1$-$m3$, the ITMRs are fixed at $1:10$ with different inclination angles, while the mass ratio in models $m4$-$m6$ are $3:10$. The projected density profiles when rings propagate out to about $3h$ are shown in Fig. \ref{den2d_mt}, together with $t1$-$t3$ models whose mass ratios are $1:2$ for a comparison. We find that the clumpy models can also produce a ring-like structure with a low ITMR of $1:10$. But in the centres of the ring galaxies, there are massive clumps and bridge structures connecting the centres and the rings. The central structures disappear when ITMR increases to $3:10$. The rings tend to be stronger with the increase of ITMR. When ITMR is fixed at a certain value, the ring-like structures becomes more elliptical with the raising inclination angle.

To summarise, a knotty ring-like structure emerges in simulations with different orbital parameters, including different inclination angles, initial relative velocities and ITMRs. The clumpy galaxies can lead to a ring-like structure if $b$ takes small values, while large values of $b$ result in a lopsided arc structure.
For a low ITMR of $1:10$, central dense structures are generated in the target galaxy models. The central structures disappear as the value of ITMR increases. A thick, knotty and empty ring structure is formed in most of the collisions on the head-on and off-centre orbits. Since galaxies are more clumpy (unstable) in the early universe with $1<z<5$ \citep{Conselice+2005} and interactions among galaxies are more frequent \citep{Patton+2002}, our results indicate that the empty ring galaxies should be observed at high redshifts. The observations of a knotty empty ring galaxy at $z=2.19$ \citep{Yuan+2020} can be viewed as a strong support to our theoretical predictions.

\section{The existing timescales of the clumps undergoing collisions}\label{sec:ts}
\begin{figure}
\centering
\includegraphics[width=90mm]{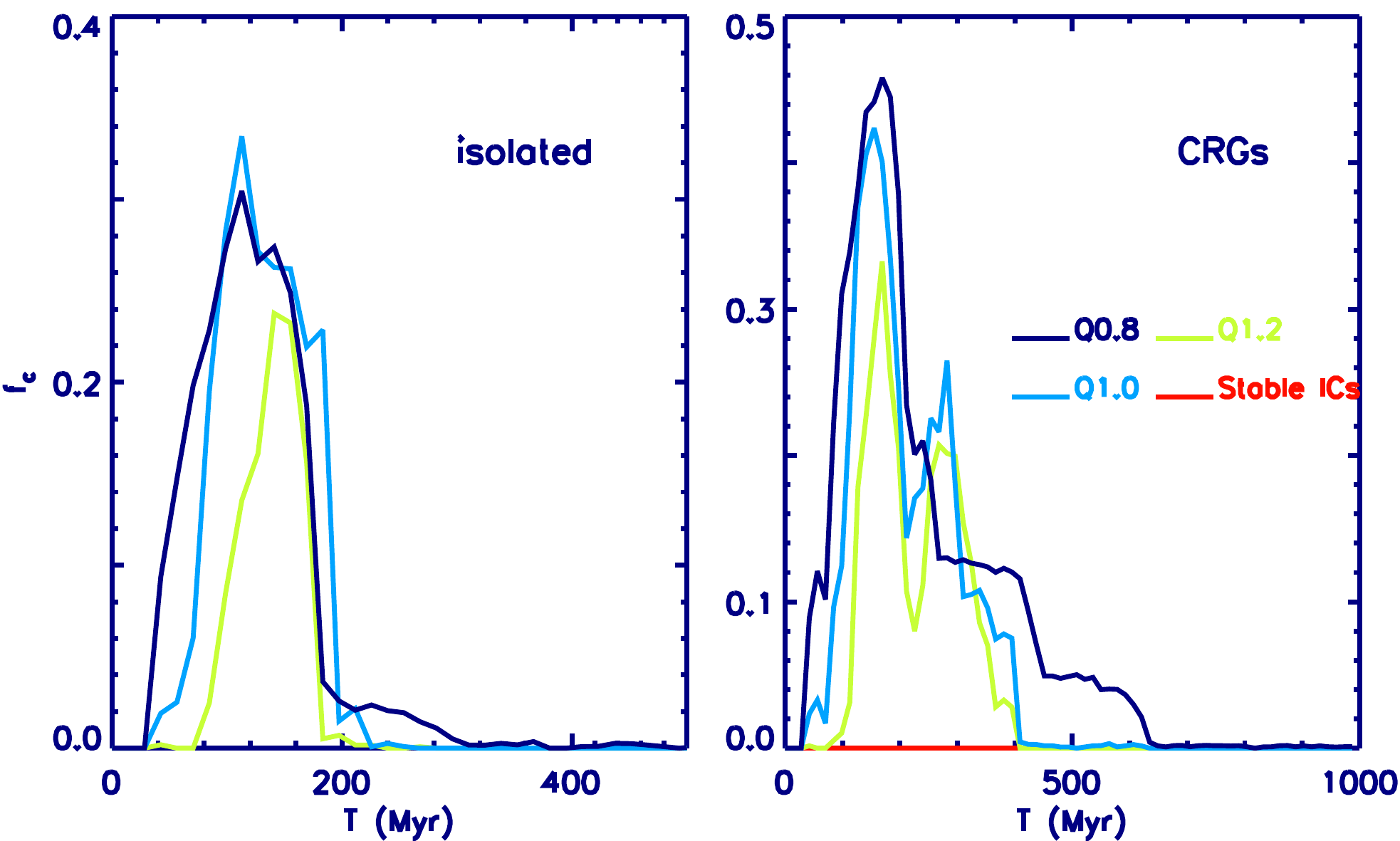}
\caption{The fraction between clumps-and-overall-stellar masses, $\fc$, for the self-evolved unstable discs (left panel) and for the unstable discs collided with a dwarf galaxy (right panel). There is a comparison collision between a stable disc model and a dwarf on the right panel, in which no clumps form within 1 Gyr.}\label{timescale}
\end{figure}

As the clumps in a galaxy spread outwards together with the propagation of the ring structure after the collision, the background disc density where the clumps are located is reduced as the galactocentric radius grows. The dynamical friction between the clumps and the disc plane becomes weaker when they move to the outer region of a galaxy. In an isolated clumpy galaxy, the clumps evolve near the radius where they form, normally within a few times of the scale length, $h$, of the disc and therefore the background disc density does not change significantly. The dynamical friction between the clumps and the disc plane of an isolated galaxy is expected to be larger than that of a collisional galaxy, and the existing timescales of the clumps are expected to be longer in the later case.

The fractions between clumps and overall stellar masses, $\fc$, for the clumpy galay models have been presented in the isolated (left panel) and in the collisional (right panel) simulations in Fig. \ref{timescale}. For the isolated $q1$-$q3$ models, the clumps form at the early stage of the self-evolution of galaxies and the values of $\fc$ reach the maximal within $100-150 ~\Myr$ and decrease down to less than $0.02$ after $300 \Myr$. The peak values of $\fc$ are within the range of $[0.24,~0.34]$. Those results agree well with some previous studies on the formation of clumps through instability \citep[e.g.,][]{Bournaud+2007,Elemgreen+2008}. The existing timescales of the clumps are slightly shorter than that obtained by \citet{Bournaud+2007} and in \citet{Elemgreen+2008}, since our density distribution of the disc ICs are exponential while theirs are uniform. The disc density is higher in smaller radii of our models, and the clumps vanish a little more rapidly due to the stronger dynamical friction from the disc plane.

On the other hand, the clumpy structures propagate radially after a minor collision. Due to the lower disc density distribution on the outwards orbits of the clumps, the dynamical friction from the disc plane is too weak to digest those clumps rapidly. The clumpy state of the collided galaxy therefore exists for approximately $0.4-0.6$ Gyr for the $q1$-$q3$ models, which is a factor of two longer compared to those of isolated self-evolved clumpy galaxies. 
Moreover, the peak values of $\fc$ rang from 0.34 to 0.45 for the collision models of $q1$-$q3$, which are significantly larger than those of the isolated models. The larger mass fraction of clumps in the collision models results from the gravitational instability caused by the perturbation of intruder galaxy.
The values of $\fc$ become smaller and the existing timescale gets shorter with the increase of $\Q0$ of the model after a collision, since models are becoming more stable. Thus, a collision with a larger $q0$ generates a much smoother ring structure instead of clumps (see Fig. \ref{den2d_q}). 

Furthermore, we have calculated the $\fc$ for models $m1$-$m6$ and $t1$-$t3$. The peak values of $\fc$ are approximately $0.33$ for models with ITMR of $0.1$, $0.35-0.42$ for models with ITMR of $0.3$, and $0.44$ for models with ITMR of $0.5$. In general, the mass of the clumps increases as the ITMR grows. With a fixed value of ITMR, the change of inclination angle does not introduce additional formation of clumps. The existing timescale of the clumpy-state in the collisions become longer when the ITMR is larger. In the minor collisions with ITMR of $0.1$ (i.e., models $m1$-$m3$), the existing timescales are approximately $300~\Myr$ for the models, which is almost the same as that of an isolated self-evolved model. Thus a minor collision with ITMR of $0.1$ barely affects the formation and disruption of the clumps.

\section{Conclusions and discussions}\label{sec:conclusion}
We have systematically studied the structure formation and evolution of a clumpy-state galaxy collide with an companion galaxy on various orbits, including head-on collisions and off-centre collisions. We have shown that an empty collisional ring galaxy can be naturally formed in the early universe when the disc progenitors are still very young and clumpy, without the necessity of introducing strict conditions, such as a special inclination angle for the orbit of the intruder galaxy or a special projection angle for the target galaxies on the light-of-sight. For the off-centre collisions, a ring forms in most of the orbits with different values of inclination angle, initial relative velocity, ITMR and also of small values of impact factor. A galaxy pair on an orbit with a large value of impact factor leads to an arc structure instead of a complete ring. In observations, the number density of galaxies at $z\approx 2$ is about 10 times larger than that in the local universe \citep{Yuan+2020}, thus collisions or mergers between galaxies should be more frequent at such a high redshift. The results in our simulations provides another natural explanation for the observed empty ring at high redshift \citep{Yuan+2020}.

Besides the prediction of empty GRGs in the early universe, our simulations lead to an additional prediction. According to our results, the clumps exist for a longer timescale if the galaxy undergoes a collision. A longer existing timescale of the clumps in the collisions of galaxies indicates a higher number density of the clumpy (face-on) or chain (edge-on) state of galaxies at higher redshifts when galaxies frequently interact with each other.

In our simulations, we ignore the gas dynamics. The dynamical evolution of stellar components in a pure gravitational system and in a mixture system including stars and gas are very similar. The gas dynamics do not significantly change the morphology of a knotty ring-like structure for the target galaxy in a collision. We shall systematically study the evolution of the collisional rings in the early universe including star formation and thermodynamics in a subsequent project.

\section{Acknowledgments}
XW thanks for support through grants from the Natural Science Foundation of China (Number NSFC-12073026, NSFC-11421303) and ``the Fundamental Research Funds for the Central Universities''.

XW designed and supervised the project. BG performed the simulations with assistance of XW and GC. BG and XW analyzed the results. XW wrote the manuscript with contributions from BG and GC.

\bibliographystyle{aasjournal}
\bibliography{ref}

\begin{appendix}

\begin{table}
\centering
\caption{
The grid resolution parameters for the self-evolved model, $q1$. The maximal allowed levels of refinement and the actual maximal levels of refinement are listed in the $2^{nd}-3^{rd}$ columns. At each coarse time step, an AMR cell is further divided into eight subcells if there are more than $n_{\rm refine}$ particles in a cell. The actual grid resolution, $\Delta x^{\rm actual}$, are shown in the $5_{th}$ column. The numbers of stellar and dark matter halos are listed in the $6_{th}$ through $6_{th}$ columns.
}\vskip 0.5cm
\begin{tabular}{llllllllllllll}
\hline
Resolution  & $l_{\rm max}$ &$l_{\rm max}^{\rm actual}$ & $n_{\rm refine}$ & $\Delta x^{\rm actual}$& $N_*$ & $N_{\rm halo}$ \\
tests & & & & (pc) & $10^5$  & $10^5$ \\
\hline
r1 & 10 & 10 & 10 & 488.3  & 13.0 & 10.0 \\
r2 & 12 & 12 & 10 & 122.1  & 13.0 & 10.0 \\
r3 & 14 & 14 & 10 & 30.5   & 13.0 & 10.0 \\
r4 & 16 & 16 & 10 & 7.6  & 13.0 & 10.0 \\
r5 & 24 & 17 & 5  & 3.8  & 13.0 & 10.0 \\
\hline
\end{tabular}
 \label{orbitres}
\end{table} 	

\begin{figure*}
\centering
\includegraphics[width=160mm]{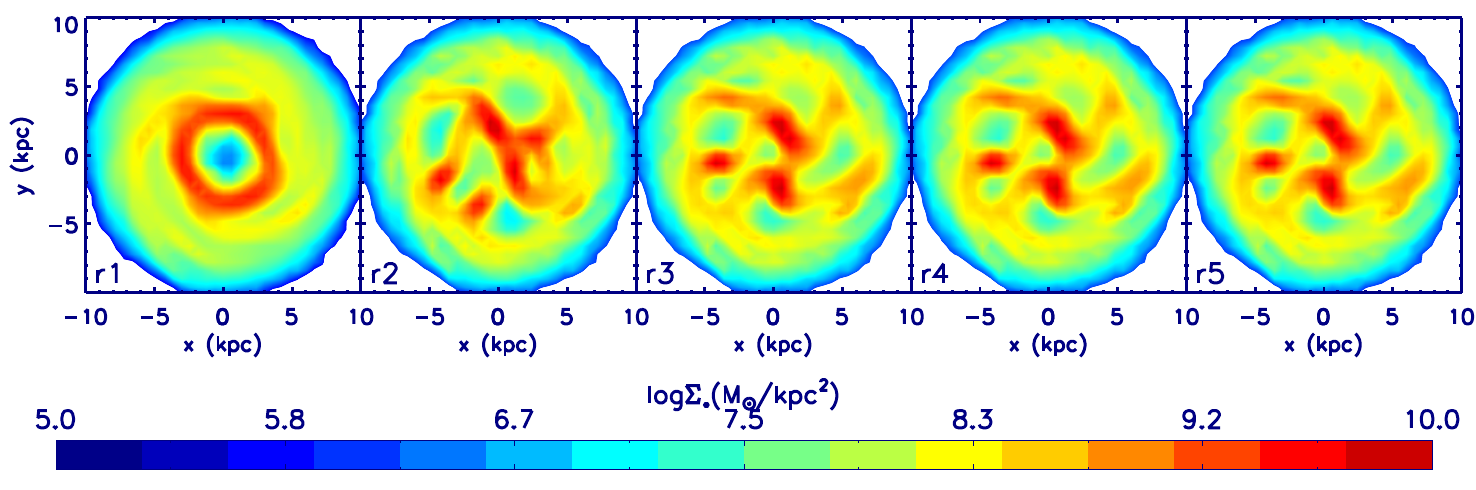}
\caption{Projected mass density profiles of the self-evolved model $q1$ with different grid resolution.}\label{den2d_res}
\end{figure*}

\section{Tests for gravitational instability amplified by numerical noise}\label{sec:resolution}
In the simulations by \citet{Tamburello+2015}, the fragmentation of a disc is suppressed with increasing grid resolution. The sizes and numbers of the clumps in a galaxy are over-predicted in a simulation with low grid resolution. Here we provide a set of simulations testing the grid resolution. We choose the model parameters of $q1$ in Table \ref{orbit} for the ICs. There are five simulations performed for these ICs with different choices of grid resolution or different numbers of particles. The resolution parameters are listed in Table \ref{orbitres}. The ICs are self-evolved for $100~\Myr$, at which the mass fraction of clumps in a galaxy model reaches the maximal value. The clumpy states of the models with different grid resolutions are shown in Fig. \ref{den2d_res}. 

For the model with minimal grid resolution, i.e., model $r1$, whose actual spatial resolution is $\Delta x^{\rm actual}=488$ pc, the structure is too ambiguous to distinguish the clumps. In fact, there is only a ring-like structure after self-evolving for $100~\Myr$. Such a structure is unreliable since the spatial resolution is too low. With the increase of $l_{\rm max}$, the clumpy structures appear in models $r2$-$r5$. In model $r2$, with a lower spatial resolution of $\Delta x^{\rm actual}=122$ pc, the sizes of clumps are larger than those in models $r3$-$r5$, which agrees well with the results in \citet{Tamburello+2015}. However, keeping the number of particles unchanged, when the spatial resolution reaches $\Delta x^{\rm actual}\leq 31$ pc, the numbers and sizes of the clumps are indistinguishable from the projected density. Since the actual spatial resolution in our simulations is approximately $10$ pc (see Sec. \ref{sec:nbody}), the artificial fragmentation from particle noise is avoided in our models. We further examine the mass fraction of the clumps, $f_{\rm C}$, after the models have self-evolved for $100~\Myr$. For models $r2$-$r5$, $f_{\rm C} \in [0.26,~0.28]$. The different values of spatial resolution do not lead to distinct overall mass of clumps.

\begin{table*}
\centering
\caption{
The model parameters of the collisions between the unstable target galaxy and the intruder galaxy. The names of the orbits and the ITMR are listed in the $1^{st}-2^{nd}$ columns. The Toomre instability parameter for the stellar component ($Q_*$) and for the overall baryonic matter ($\Q0$) of the target galaxies are given in the $3^{rd}-4^{th}$ columns. The initial positions of the intruder galaxy in galactocentric coordinates of the targets are shown in the $5^{th}$ column. The $6^{th}$ column lists the impact parameter, $(b)$. The initial relative velocity, $v_{\rm 0}$, and its x$-$ and z$-$axial components, $v_{\rm 0x}$ and $v_{\rm 0z}$ are shown in the $7^{th}-9^{th}$ columns. The $10^{th}$ column is the inclination angle, $\theta$. The $11^{th}$ through $13^{th}$ columns represent the masses of the stellar and gas component of the target galaxy models, and the number of particles in the simulations.
}\vskip 0.5cm
\begin{tabular}{llllllllllllll}
\hline
Collisional  & $M_{\rm i}/M_{\rm t}$ & $Q_*$& $Q_0$& $(x_0,~y_0,z_0)$ & $b$ &$v_{\rm 0}$ & $v_{\rm 0x}$ & $v_{\rm 0z}$ & $\theta$ &$M_*$  &$M_{\rm{gas}}$&$N$ \\
models & &  & (kpc) & (kpc) & (km/s) & (km/s) & (km/s) & ($^\circ$)   & ($10^{10}\Msun$)& ($10^{10}\Msun$) & -\\
\hline
g1 &0.5& 1.5&0.8 & $(0.0,~0.0,~-50.0)$ &0.0&600& 0 & 600 & 0.0 & 3.0 & 3.0 & $1.0\times 10^6$ \\
g2 &0.5& 1.5&0.8 & $(0.0,~0.0,~-50.0)$ &0.0&600& 0 & 600 & 0.0 & 3.0 & 3.0 & $6.7\times 10^5$ \\
g3 &0.5&0.8&0.8 & $(0.0,~0.0,~-50.0)$ &0.0&600& 0 & 600 & 0.0 & 6.0 & 0.0 & $6.3\times 10^5$ \\
\hline
\end{tabular}
 \label{orbithydro}
\end{table*}

\section{The self-evolution of clumpy galaxies with and without hydrodynamics}\label{sec:hydro}
\begin{figure*}
\centering
\includegraphics[width=160mm]{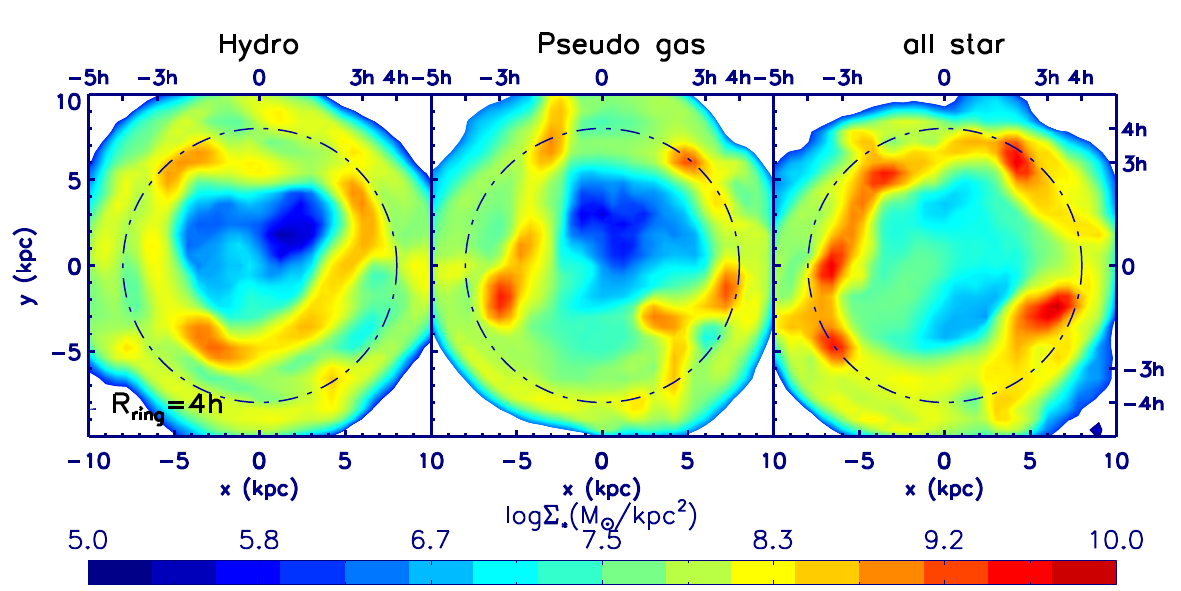}
\caption{Projected mass density profiles of the collisional rings propagating to the radii of 3h of the target galaxy for the same mass distributed models with (left panel) and without (middle panel) hydrodynamics. There is a comparison of model with only stellar component in the right panel.}\label{den2d_gas}
\end{figure*} 

We set up a set of simulations to test how the gas affects the formation and evolution of the collisional rings in the early universe. The orbital parameters of simulated models are listed in Tab. \ref{orbithydro}. The target galaxy of model $g1$ includes a stellar disc, a gaseous disc and a dark matter halo. The mass distribution of stellar component is the same as that of model $q1$, but the stellar mass is reduced to $3\times 10^{10}\msun$. The mass distribution of gas follows the density profile of the stellar component in \S \ref{sec:models} and the mass of gas is also $3\times 10^{10}\msun$. Therefore, the overall mass of baryonic matter remains $M_{\rm d}$. The temperature and the velocity dispersion of gas are $10^4~K$ and $50~\kms$, respectively. The Toomre parameter, $Q_*$ is $1.5$ for the stellar component, and $\Q0$ is $0.8$ and for the overall of baryonic matter, which is defined by
\beq Q_0 (r) \approx \left( \frac{\pi G \Sigma_{\rm gas}}{\sigma_{\rm gas}\kappa(r)}+\frac{1}{Q_*}\right)^{-1}.\eeq
Here $\sigma_{\rm gas}$ is the velocity dispersion of gas.

The simulation of model $g1$ is performed via {\it RAMSES}, with a box size of $2500~\kpc$ and a maximal refinement level of $14$, which corresponds to a grid resolution of $2500/2^{14}~\kpc\approx 0.15~\kpc$ for gravitational potential.
The mass distribution of model $g2$ is the same as that of model $g1$, but the gas is simulated by considering only the gravitation. Thus the gas component is pseudo-gas, i.e., providing only a gravitational potential. The hydrodynamics is ignored in $g2$ model. The baryonic matter only contains a stellar component in $g3$ model. The overall baryonic mass of models $g2$ and $g3$ are exactly the same. All the three collisional models are simulated on a head-on orbit. After a collision with an ITMR of $0.5$, there appears a knotty stellar ring structure in all the three models (Fig. \ref{den2d_gas}). All the galaxies exhibit empty ring structures. For the $g1-g2$ models, the knotty ring-like stellar structures appear to be very similar to each other irrespective of whether hydrodynamics is taken into account. The ring is the strongest in the $g3$ model, since the stellar mass is a factor of two larger than the other two models. In general, the evolution of stellar structure for galaxies undergoing collision are very similar with and without hydrodynamics. Thus we mainly focus on the pure gravitational evolution of CRGs formed from clumpy-state galaxies in this work. A more systemic study on gas dynamics, star formation and etc will be performed in a future project.

\end{appendix}
\end{document}